\documentclass[preprint,aps,prl,showpacs,psfig,onecolumn]{revtex4}

\usepackage{amssymb}
\usepackage{amsmath}
\usepackage{graphicx}

\begin{document}

\title{Single or multi-flavor Kondo effect in graphene}

\author{Zhen-Gang Zhu$^{1}$, Kai-He Ding$^{2}$, Jamal Berakdar$^{1}$}
%
\address{$^{1}$IInstitut f\"{u}r Physik, Martin-Luther-Universit\"{a}t
Halle-Wittenberg, Nanotechnikum-Weinberg, Heinrich-Damerow-St. 4,
06120 Halle, Germany;\\
$^{2}$ Department of Physics and Electronic Science, Changsha
University of Science and Technology, Changsha 410076, China}


\begin{abstract} Based on the tight-binding formalism, we
investigate the Anderson and the Kondo model for an adaom magnetic
impurity above graphene. Different impurity positions are
analyzed. Employing a partial wave representation we study the
nature of the coupling between the impurity and the conducting
electrons. The components from the two Dirac points are mixed
while  interacting with the impurity. Two configurations are
considered explicitly: the adatom is above one atom (ADA), the
other case is the adatom above the center the honeycomb (ADC). For
ADA  the impurity is coupled with one flavor for both A and B
sublattice and both Dirac points. For  ADC the impurity  couples
with  multi-flavor states for a spinor state of the impurity. We
show, explicitly for a 3d magnetic atom, $d_{z^{2}}$,
($d_{xz}$,$d_{yz}$), and ($d_{x^{2}-y^{2}}$,$d_{xy}$) couple
respectively with the $\Gamma_{1}$,  $\Gamma_{5} (E_{1})$, and
$\Gamma_{6} (E_{2})$ representations (reps) of $C_{6v}$ group in
ADC case. The basses for these reps of graphene are also derived
explicitly. For ADA we calculate the Kondo temperature.

\end{abstract}
\maketitle


\section{Introduction}
In Graphene, a monolayer of carbon atoms
recently fabricated successfully \cite{graphene}, the valence and
the conduction bands touch at two inequivalent Dirac points $K_-$
and $ K_+$ at the corners of the first Brillouin zone (FBZ). Near
$K_-$ and $ K_+$ the low energy dispersion is linear, indicating a
massless Dirac fermions behavior. This particular  band
structure is at the heart of  a number of unusual electronic
properties \cite{neto}.
Graphene is also an
interesting candidate for transport-applications, in particular
for spintronics: It exhibits a remarkably high mobility and the
carrier density is controllable by a gate voltage; the mean free
path can be as large as 1 $\mu m$. Graphene is however not immune
to disorder that influences its electronic properties \cite{neto}.
Extrinsic disorder is realized in a variety of ways: adatoms,
vacancies, charges on top of graphene or in the substrates, and
extended defects such as cracks and edges. When  localized
magnetic impurities are added
\cite{meyer} the Kondo effect, i.e. the dynamic screening of the
localized  moment, has to be addressed at temperatures $T$ below
the Kondo temperature $T_K$. In this context, previous studies
addressed the influence of   magnetic impurities using the
Hatree-Fock approximation \cite{uchoa,ding} which is valid at
$T>T_K$ (see also  \cite{uchoa0906}). In Ref.
\cite{hentschel} the anisotropic \emph{single} channel
Kondo model  is investigated briefly and in Ref.
\cite{dora} the \emph{infinite-U} Anderson model for an
impurity embedded in a  graphene sheet has been employed and
concluded a Fermi liquid behaviour. Recently, it has been  claimed
that a two-channel Kondo in graphene is present due to the
\emph{valley degeneracy }of the Dirac electrons \cite{sengupta},
which leads to an over-screening and thus to a
non-Fermi-liquid-like ground state. Very recently Refs.
\cite{cornaglia,zhuang}  reported on  scanning tunneling
spectroscopy studies to investigate the Kondo effect in graphene.
\emph{All} of these  studies consider  a particular configuration
of the impurity. In this work, we show that the position of the
impurity on or in graphene  plays a subtle role and  affects
essentially the underlying physics. In addition to  the role of
the vanishing density of state (DOS) at the Dirac points and the
linear spectrum in their vicinity, a further important  issue  is
to clarify  whether  a single or a multi-channel Kondo problem is
realized. This is insofar important as the ground states for these
two cases are essentially different: As established
\cite{nozieres}, when the channel number $N_{\text{cha}}$ is equal
to $2S_{\text{im}}$ ($S_{\text{im}}$ is the spin of the impurity),
the impurity spin  is then compensated by the host electrons
completely, resulting in a singlet Fermi liquid ground state. When
$N_{\text{cha}}>2S_{\text{im}}$ we enter the  over-compensated
regime where an opposite spin to the original spin  appears and
acts as a remainder spin;  an effective antiferromagnetic coupling
between this remainder effective impurity spin and the conducting
spins  results in a non-Fermi liquid ground state
\cite{cox,cox96,varma}.  Thus, it is crucial to clarify when
graphene with an impurity  with a spin one-halve is a single or a
two channels Kondo  system.
Recently, scanning tunneling conductance spectra are investigated with respect to
the position  of the scanning tip  in Refs. \cite{uchoa09,saha} (see also Ref. \cite{wehling}).\\
As a first essential step we identify the physical origin of the
multi channels for the Kondo effect by analyzing the tight-binding Anderson
model \cite{anderson,hewson}. A partial wave method is  then employed to
reduce the redundant flavors and to identify the coupling flavors. The role of
the Dirac points is exposed. An adatom above
one carbon atom (ADA)  or an adatom above the center of the honeycomb (ADC)
configurations are considered.

\begin{figure}
\includegraphics[width=0.7\textwidth]{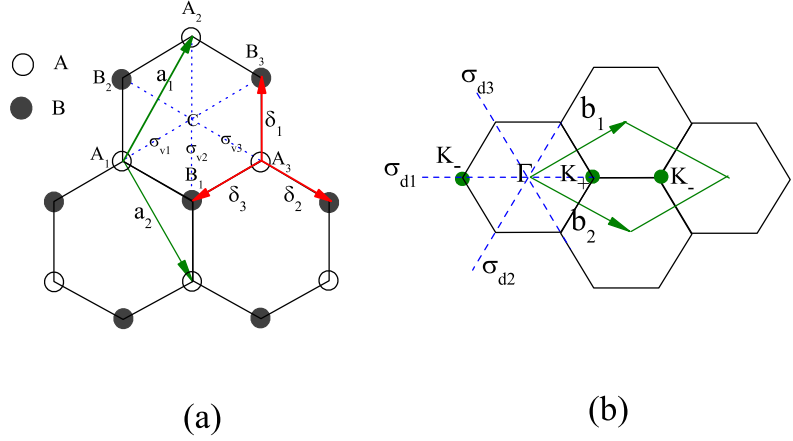}
\caption{(color online).  (a)  shows  graphene  primitive vectors and the symmetry
operations defining the group of the crystal, $\sigma_{vi},i=1,2,3$ are reflection planes showing
up in $C_{6v}$ little group at $\Gamma$ point.  (b) shows  FBZ   with the symmetry operations
(see text for explanations). \label{figpotential}}
\end{figure}

\section{Theoretical formulation}
We start from the Anderson Hamiltonian
\begin{equation}
H=H_{\text{g}}+H_{f}+H_{\text{hyb}}. \label{totalh1}
\end{equation}
 $H_{\text{g}}$ ($H_{f}$) describes graphene,
(impurity)  and $H_{\text{hyb}}$ stands for the impurity-graphene hybridization.
%
The graphene primitive vectors  are
$\mathbf{a}_{1}=a(\frac{1}{2},\frac{\sqrt{3}}{2})$,
$\mathbf{a}_{2}=a(\frac{1}{2},-\frac{\sqrt{3}}{2})$, where
$a=\sqrt{3}a_{\text{cc}}$, $a_{\text{cc}}$ is the distance of the
nearest carbon atoms. The B sublattice is related to A sublattice
by $\boldsymbol{\delta}_{1}=(\mathbf{a}_{1}-\mathbf{a}_{2})/3$,
$\boldsymbol{\delta}_{2}=\mathbf{a}_{1}/3+2\mathbf{a}_{2}/3$, and
$\boldsymbol{\delta}_{3}=-\boldsymbol{\delta}_{1}-\boldsymbol{\delta}_{2}=-2\mathbf{a}_{1}/3-\mathbf{a}_{2}/3$
(cf.~Fig. \ref{figpotential}(a)). The two inequivalent Dirac points
in FBZ are $\mathbf{K}_{\pm}=\pm\frac{2\pi}{a}(2/3,0)$. The second
quantization tight-binding-Hamiltonian
\cite{neto,uchoa,semenoff,gusynin,divincenzo} reads  $
H_{\text{g}}=\sum_{\langle
ij\rangle\sigma}(ta^{\dagger}_{i\sigma}b_{j\sigma}+
t^{*}b^{\dagger}_{j\sigma}a_{i\sigma}) $ where  the sum  runs over
 the nearest-neighbor pairs  $\langle ij\rangle$, $\sigma$ is a spin index,
and $a_{i\sigma}(b_{i\sigma})$ are
annihilation  operators for states on the A (B) sublattice.
The Hamiltonian in
momentum space is
$H_{\text{g}}=\sum_{\mathbf{\bar{k}}\sigma}[\xi(\mathbf{\bar{k}})
a^{\dagger}_{\mathbf{\bar{k}}\sigma}b_{\mathbf{\bar{k}}\sigma}+\xi^{*}(\mathbf{\bar{k}})b^{\dagger}_{\mathbf{\bar{k}}\sigma}a_{\mathbf{\bar{k}}\sigma}],
$
where $\xi(\mathbf{\bar{k}})=-t\Phi(\mathbf{\bar{k}})$,  $\Phi(\mathbf{\bar{k}})=\sum_{\boldsymbol{\delta}_{i}}e^{i\mathbf{\bar{k}}\cdot\boldsymbol\delta_{i}}
$, and $\mathbf{\bar{k}}$ are wave vectors in FBZ. 
The hybridization Hamiltonian in tight-binding formalism is
$H_{\text{hyb}}=\sum_{\mathbf{\bar{k}}\sigma}[(V^{Af}_{\mathbf{\bar{k}}}a^{\dagger}_{\mathbf{\bar{k}}\sigma}f_{\sigma}+
V^{Bf}_{\mathbf{\bar{k}}}b^{\dagger}_{\mathbf{\bar{k}}\sigma}f_{\sigma})+h.c.],$
where
$V^{\alpha f}_{\mathbf{\bar{k}}}=\frac{1}{\sqrt{N}}\sum_{\mathbf{R}_{i}\neq\mathbf{R_{\text{im}}}}e^{-i\mathbf{\bar{k}}\cdot(\mathbf{R}_{i}+\boldsymbol{\tau}_{\alpha})}
V^{\alpha f}(\mathbf{R}_{i}+\boldsymbol{\tau}_{\alpha}-\mathbf{R}_{\text{im}})$, $\alpha$=A or B, $\boldsymbol{\tau}_{\alpha}$
stands for the relative location for $\alpha$ atom in the unit cell,
$V^{\alpha f}(\mathbf{R}_{i}+\boldsymbol{\tau}_{\alpha}-\mathbf{R}_{\text{im}})=\int d\mathbf{r}[\phi^{\alpha}(\mathbf{r}-\mathbf{R}_{i}-\boldsymbol{\tau}_{\alpha})]^{*}h(\mathbf{r})\phi_{L}(\mathbf{r}-\mathbf{R}_{\text{im}}),$
and $\phi^{\alpha}$ is the atomic function for the atoms in $\alpha$ sublattice (only  $\pi$ orbitals are included),
$\phi_{L}$ is the localized impurity wave function,  $h$ is the single particle Hamiltonian, and $N$ is the number
of unit cells.

$V^{\alpha f}(\mathbf{R}_{i}+\boldsymbol{\tau}_{\alpha}-\mathbf{R}_{\text{im}})$ is a Slater-type bond in LCAO formalism \cite{slater}
which describes the strength of hybridization between two atomic orbitals located at the impurity and its neighbors.
Here only $p_{\pi}$ electrons are relevant. They  form  the $\pi$ and $\pi^{*}$  which touch at
the Dirac points $\mathbf{K}_{\pm}$.
An s-wave impurity substituting one carbon atom in the graphene plane
is decoupled from $\pi$ and $\pi^{*}$ bands;
for d- and f-wave impurity in this substitution case is possible.
Our focus  is however on  the case where the impurity is above the graphene plane.
The relative positions of the impurity
determine the hybridization strength, i.e. $V^{\alpha f}$ and the phase is included in the
exponential in $V^{\alpha f}_{\mathbf{\bar{k}}}$. 

%
%
To obtain an effective low-energy Hamiltonian  we expand the wave vector $\mathbf{\bar{k}}=\mathbf{K}_{\pm}+\mathbf{k}$ around $\mathbf{K}_{\pm}$.
Thus, $\Phi(\mathbf{\bar{k}})\approx(-v_{F}/t)(\pm k_{x}-ik_{y})$, or
$ \Phi(\mathbf{\bar{k}})|_{s}\approx(-v_{F}/t)k\, \lambda e^{-i\lambda\theta},$
where $\lambda=\text{sgn}(s)=1(-1)$ for $s=K_{+}(K_{-})$, $v_{F}$ is  graphene Fermi's velocity, $\theta$
is the azimuthal angle of $\mathbf{k}$.
The Hamiltonian of graphene is then
\begin{equation}
H_{\text{g}}=\sum_{s\sigma\mathbf{k}}v_{F}k\left(a_{s\mathbf{k}\sigma}^{\dagger},b_{s\mathbf{k}\sigma}^{\dagger}\right)
\left(
\begin{array}{cc}
0 & \lambda e^{-i\lambda\theta} \\
\lambda e^{i\lambda\theta} & 0
\end{array}
\right)
\left(
\begin{array}{c}
a_{s\mathbf{k}\sigma} \\
b_{s\mathbf{k}\sigma}
\end{array}
\right). \label{hg}
\end{equation}
 $H_{\text{g}}$ is diagonalized by  introducing the $\zeta$ fields \cite{cassanello} as
\begin{eqnarray}
a_{s\mathbf{k}\sigma} &=& \frac{1}{\sqrt{2}}[\zeta_{+\sigma}^{s}(\mathbf{k})+\zeta_{-\sigma}^{s}(\mathbf{k})],\: 
b_{s\mathbf{k}\sigma} = \frac{\lambda}{\sqrt{2}} e^{i \lambda\theta}[\zeta_{+\sigma}^{s}(\mathbf{k})-\zeta_{-\sigma}^{s}(\mathbf{k})],
\label{abfield}\\
H_{\text{g}}&=&\sum_{s\sigma\mathbf{k}}\left[v_{F}|k|\zeta_{+\sigma}^{s\dagger}(\mathbf{k})\zeta_{+\sigma}^{s}(\mathbf{k})
-v_{F}|k|\zeta_{-\sigma}^{s\dagger}(\mathbf{k})\zeta_{-\sigma}^{s}(\mathbf{k})\right].
   \nonumber
\end{eqnarray}

\subsection{Partial wave method and hybridization Hamiltonian}
Going over from a discrete 
to a continuum $\mathbf{k}$  \cite{krishna} we write
$
H_{\text{g}} =\sum_{s\sigma}\int d\mathbf{k}\left[v_{F}|k|\zeta_{+\sigma}^{s\dagger}(\mathbf{k})\zeta_{+\sigma}^{s}(\mathbf{k})
-v_{F}|k|\zeta_{-\sigma}^{s\dagger}(\mathbf{k})\zeta_{-\sigma}^{s}(\mathbf{k})\right],$ and $
H_{\text{hyb}} = \frac{(N\Omega_{0})^{1/2}}{2\pi}\sum_{s\sigma}\int d\mathbf{k}\left[\left(V_{s\mathbf{k}}^{Af}a_{s\mathbf{k}\sigma}^{\dagger}f_{\sigma}+
V_{s\mathbf{k}}^{Bf}b_{s\mathbf{k}\sigma}^{\dagger}f_{\sigma}\right)+h.c.\right].
$
 $\Omega_{0}$ is the   unit cell area.
In
 2D orbital momentum eigenfunctions $e^{im\theta}$   we write
$\zeta_{\pm\sigma}^{s}(\mathbf{k})=\frac{1}{\sqrt{|k|}}\sum_{m=-\infty}^{\infty}\frac{1}{\sqrt{2\pi}}e^{im\theta}\zeta_{\pm\sigma}^{ms}(k),
$
where $\{\zeta_{\upsilon\sigma}^{ms}(k),\zeta_{\upsilon'\sigma'}^{m's'\dagger}(k')\}=\delta_{ss'}\delta_{\upsilon\upsilon'}\delta_{\sigma\sigma'}\delta_{mm'}\delta(k-k')$.
$H_{\text{g}} $ is then expressed as
\begin{equation}
H_{\text{g}} =\sum_{ms\sigma}\int_{0}^{\infty} v_{F}|k|dk\left[\zeta_{+\sigma}^{ms\dagger}(k)\zeta_{+\sigma}^{ms}(k)
-\zeta_{-\sigma}^{ms\dagger}(k)\zeta_{-\sigma}^{ms}(k)\right].
\end{equation}
For  discussing the hybridization Hamiltonian for various geometric configurations we set the spatial
zero point "O" as the projection of impurity position onto the unit cell.\\
\emph{Adatom impurity above one A atom (ADA)}:
In this case $\tau_{A}=0$, in the Dirac cones we find  (we assume $v_{0}$ is real)
$V^{Af}_{s\mathbf{k}}=\frac{V^{Af}(0)}{\sqrt{N}}=\frac{v_{0}}{\sqrt{N}}.$
The next nearest neighbors are three B atoms
at $\boldsymbol{\delta}_{i}, i=1,2,3$. Thus,
$V^{Bf}_{s\mathbf{k}}=\frac{\Phi^{*}(\mathbf{k})V^{Bf}}{\sqrt{N}}=\lambda\frac{v_{F}kv_{1}e^{i\lambda\theta}}{-t\sqrt{N}},
$
where $v_{1}=V^{Bf}$. Substituting in  $H_{\text{hyb}}$  we find
$$H_{\text{hyb}} =\sum_{s\sigma}\int_{0}^{\infty}\frac{\sqrt{\pi\Omega_{0}|k|}dk}{2\pi}
\left[v_{0}(\zeta_{+\sigma}^{0,s\dagger}(k)+
\zeta_{-\sigma}^{0,s\dagger}(k))
+ \frac{v_{1}v_{F}k}{-t}(\zeta_{+\sigma}^{0,s\dagger}(k)-
\zeta_{-\sigma}^{0,s\dagger}(k))\right]f_{\sigma}+h.c. .
$$
This equation tells  us only $m=0$ flavor  couples to the impurity. \\
\emph{Adatom impurity above the center of the honeycomb}:
 Here, the impurity  hybridizes with the same  strength  $v_{0}$, with  6 nearest neighbors, 3 A atoms
and 3 B atoms. With respect to O,
$A_{2}, B_{2}, A_{3}, B_{3}$ are  in other unit cells. 
Three A(B) atoms yield $\Phi^{*}(\Phi)$ when the phases are coherently superimposed, i.e.
$
V^{Af}_{\mathbf{k}}a_{\mathbf{k}\sigma}^{\dagger}f_{\sigma}$+$V^{Bf}_{\mathbf{k}}b_{\mathbf{k}\sigma}^{\dagger}f_{\sigma}
= $\\ $\frac{v_{0}}{\sqrt{N}}
\left[
e^{-i\mathbf{k}\cdot\boldsymbol\delta_{3}}\left(1+e^{-i\mathbf{k}\cdot(\mathbf{a}_{1}+\mathbf{a}_{2})}+e^{-i\mathbf{k}\cdot\mathbf{a}_{1}}\right)
a_{\mathbf{k}\sigma}^{\dagger}f_{\sigma}
+e^{i\mathbf{k}\cdot\boldsymbol\delta_{1}}\left(1+e^{-i\mathbf{k}\cdot\mathbf{a}_{1}}+e^{i\mathbf{k}\cdot\mathbf{a}_{2}}\right)
b_{\mathbf{k}\sigma}^{\dagger}f_{\sigma} \right]
$\\ $=\frac{v_{0}}{\sqrt{N}}\left[\Phi^{*}(\mathbf{k})a_{\mathbf{k}\sigma}^{\dagger}f_{\sigma}+\Phi(\mathbf{k})b_{\mathbf{k}\sigma}^{\dagger}f_{\sigma}
\right].$
For $H_{\text{hyb}}$ we deduce
\begin{eqnarray}
H_{\text{hyb}}&=&\sqrt{\pi\Omega_{0}}\frac{v_{0}}{-t}\sum_{s\sigma}\lambda\int_{0}^{\infty}\frac{\sqrt{|k|}dk}{2\pi}v_{F}|k|
\left(\left[(\zeta_{+\sigma}^{\lambda,s\dagger}(k)+\zeta_{-\sigma}^{\lambda,s\dagger}(k))f_{\sigma} \right.\right. \notag \\
&+& \left.\left.\lambda(\zeta_{+\sigma}^{-2\lambda,s\dagger}(k)-
\zeta_{-\sigma}^{-2\lambda,s\dagger}(k))f_{\sigma}\right] +h.c.\right).
\label{hc2}
\end{eqnarray}
When $s=K_{+}, \lambda=1$, the impurity couples with the partial wave $m=1$ from A
sublattice, and $m=-2$ from B sublattice. When $s=K_{-}, \lambda=-1$, it interacts with $m=-1$ from A
sublattice, and $m=2$ from B sublattice.\\
\section{Flavor right movers and symmetry analysis}
We unfold the range of momenta $k$ from $(0,\infty)$ to $(-\infty,+\infty)$ by defining flavor right
movers. For ADA case applies
$c_{1\sigma}^{s}(k)=\zeta_{+\sigma}^{0,s}(|k|), \text{ for } k>0 $ and $
c_{1\sigma}^{s}(k)=\zeta_{-\sigma}^{0,s}(|k|), \text{ for } k<0.$
We obtain thus
$H_{\text{g}}=\sum_{s\sigma}\int_{-\infty}^{\infty}\varepsilon_{k}dk c_{1\sigma}^{s\dagger}(k)c_{1\sigma}^{s}(k), $  $
H_{\text{hyb}}=\sqrt{\pi\Omega_{0}}\sum_{s\sigma}\left[\int_{-\infty}^{\infty}\frac{\sqrt{|k|}dk}{2\pi}\left(v_{0}+
\frac{v_{1}v_{F}k}{(-t)}\right)c_{1\sigma}^{s\dagger}(k)f_{\sigma}+h.c.\right],$
where $\varepsilon_{k}=\hbar v_{F}k$. So, the impurity  couples with only one
flavor of conducting waves.
For ADC case the impurity couples with $m=\lambda$ or $m=-2\lambda$ flavors from
A or B sublattices respectively. Therefore, we introduce two flavor right movers as
$c_{1\sigma}^{s}=\lambda\zeta_{+\sigma}^{\lambda,s}(|k|)$, $c_{2\sigma}^{s}$=$\zeta_{+\sigma}^{-2\lambda,s}(|k|)$ for $k>0$; $c_{1\sigma}^{s}$=-$\lambda\zeta_{-\sigma}^{\lambda,s}(|k|)$, $c_{2\sigma}^{s}$=$\zeta_{-\sigma}^{-2\lambda,s}(|k|)$ for $k<0$.
In terms of these, $H_{\text{g}}$=$\sum_{ns\sigma}\int_{-\infty}^{\infty}\varepsilon_{k}dk c_{n\sigma}^{s\dagger}(k)c_{n\sigma}^{s}(k)$, and $H_{\text{hyb}}$=$\frac{v_{0}\sqrt{\pi\Omega_{0}}}{(-t)}\sum_{ns\sigma}\left[\int_{-\infty}^{\infty}\frac{\varepsilon_{k}\sqrt{|k|}dk}{2\pi}
c_{n\sigma}^{s\dagger}(k)f_{\sigma}+h.c.\right]$,
where $n=1,2$. In these two cases, we have $\{c_{n\sigma}^{s}(k),c_{n'\sigma'}^{s'\dagger}(k')\}=\delta_{nn'}\delta_{ss'}\delta_{\sigma\sigma'}\delta(k-k')$.
\subsection{Symmetry analysis}
The valence bands formed by $\sigma$ bonds of $sp^{2}$ hybridization were
discussed  by Lomer long ago \cite{lomer}. The popular tight-binding Hamiltonian for
$\pi$ and $\pi^{*}$ bands formed by $p_{z}$ orbitals located on each carbon atom was developed
by Wallace \cite{wallace}, Slonczewski and Weiss \cite{slonc}.
 A detailed single group analysis was given by Bassani and Parravicini \cite{bassani}.
Group theory has
also been used \cite{basko} to analyze the Raman scattering and electron-phonon interaction.
Trigonal band structure was analyzed in terms of  graphene double group  \cite{winkler}. Here,
we ignore  $z$ plane reflection and consider the $C_{6v}$ point single group,
and  the little group at
$K_{\pm}$ as $C_{3v}$ single group.  The symmetry group elements for  $C_{3v}$ 
at Dirac
points are  $E$, $C_{3}^{+}, C_{3}^{-}$, $\sigma_{di}$, where $i=1,2,3$.
The symmetry operation $\sigma_{di}$ leaves the Dirac points  
unchanged but  interchanges A by B atoms  in real space.  $\sigma_{vi}$ acts the
opposite way.


%
%
%
\begin{table}[!htp]
\begin{center}
\begin{tabular*}{0.85\textwidth}{@{\extracolsep{\fill}}c| c c c c c c}
\hline $C_{3v}$ & E & $\bar{E}$ & $C_{3}^{+},C_{3}^{-}$,& $\bar{C}_{3}^{+},\bar{C}_{3}^{-}$ & $\sigma_{di}$ & $\bar{\sigma}_{di}$ \\
\hline
$\Gamma_{1}$ & 1 & 1 & 1 & 1 & 1 & 1  \\
$\Gamma_{2}$ & 1 & 1 & 1 & 1 & -1 & -1 \\
$\Gamma_{3}$ & 2 & 2 & -1 & -1 & 0 & 0 \\
\hline
$\Gamma_{5}$ & 1 & -1 & -1 & 1 & i & -i \\
$\Gamma_{6}$ & 1 & -1 & -1 & 1 & -i & i \\
$\Gamma_{4}$ & 2 & -2 & 1 & -1 & 0 & 0 \\
\hline
\end{tabular*}
\caption{Character table for double group $C_{3v}$ for $K_{\pm}$ point of graphene. $\Gamma_{4,5,6}$ are extra
representations (reps) and the spinor rep is $\Gamma_{4}$.}
\end{center}
\end{table}
If we use $D^{\frac{1}{2}}$ to indicate the spinor reps for a rotation group, then for $C_{3v}$,
$\Gamma_{1}\otimes D^{\frac{1}{2}}=\Gamma_{4}$. We also have the relations $\Gamma_{5}\otimes\Gamma_{4}=\Gamma_{3}$,
$\Gamma_{6}\otimes\Gamma_{4}=\Gamma_{3}$, and $\Gamma_{4}\otimes\Gamma_{4}=\Gamma_{1}+\Gamma_{2}+\Gamma_{3}$. Without
 spin and when the Fermi level $E_F$ crosses $K_{\pm}$ we have a $\Gamma_{3}$ reps. Following
  Ref. \cite{slonc} we can show the Bloch sum for the $p_{z}$ orbitals on A and B sublattices
are the basis of the representation, i.e. $\phi^{\alpha}_{\mathbf{k}}(\mathbf{r})=\frac{1}{\sqrt{N}}\sum_{\mathbf{R}_{i}}e^{-i\mathbf{k}\cdot
(\mathbf{R}_{i}+\boldsymbol{\tau}_{\alpha})}\phi_{i}^{\alpha}(\mathbf{r}-\mathbf{R}_{i}-\boldsymbol{\tau}_{\alpha})$, where $\alpha=A$ or $B$,
$\phi_{i}^{\alpha}$ representing $p_{z}$ orbital at $\mathbf{R}_{i}+\boldsymbol{\tau}_{\alpha}$ site.
When spin is taken into account and spin orbit interaction (SOI) is ignored, the degeneracy at $K_{\pm}$ is
$2\times2=4$. When SOI is considered, the reps of double group are  split into $\Gamma_{3}\otimes D^{\frac{1}{2}}=\Gamma_{4}+\Gamma_{5}+\Gamma_{6}$.
The reps $\Gamma_{4},\Gamma_{5}$, and $\Gamma_{6}$ are degenerate in absence of  SOI at $K_{\pm}$.\\
\textbf{a)} For \textit{ADA case}
 the system has a $\bar{C}_{3v}$ symmetry (different from $C_{3v}$, the character table is  shown by table 1). The  planes reflection  symmetry in $\bar{C}_{3v}$
converts the point $K_{+}$ into $K_{-}$ and vice versa. Thus, the symmetry $\bar{C}_{3v}$  mixes
the states from the two Dirac points. In fact, this group is a subgroup of $C_{6v}$ and the symmetry
reflection planes do show up in $C_{6v}$ and map Dirac points onto each others. Thus,
only one flavor  couples with the impurity. The Hamiltonian in this case shows that a basis belonging to $\Gamma_{1}$ representation of $\bar{C}_{3v}$ can be constructed from the $E$ rep of $C_{3v}$, which leads to an invariant hybridization Hamiltonian. If we consider explicitly a 3d magnetic impurity in this case, for example Mn (as realized  in ADA case \cite{duffy}), the nonvanishing coupling is only present between the $d_{z^{2}}$ orbital of the impurity and graphene since the $d_{z^{2}}$ orbital transforms according to $\Gamma_{1}$ rep.  \\
\textbf{b)} For \textit{ADC case}
the symmetry group is $C_{6v}$ where $C_{2}$, $C_{6}^{+(-)}$ and $\sigma_{vi}$ show up in this group except the symmetry operations in $C_{3v}$. In this group, $C_{2}$, $C_{6}^{+(-)}$ transform A sublattice into B sublattice, $K_{+}$ into $K_{-}$ and vice versa. $\sigma_{vi}$ keep the sublattice unchanged but exchange $K_{+}$ and $K_{-}$. Therefore, we have to construct the basis for $C_{6v}$ by using the basis for the two Dirac points in $C_{3v}$ group. From Table 2, $E$ rep in $C_{3v}$ corresponds to the $E_{1}$ and $E_{2}$ reps in $C_{6v}$. Therefore we can use the Bloch sums $\phi^{A(B)}_{K_{\pm}\mathbf{k}\sigma}$ in the $E$ rep in $C_{3v}$ to construct the reps in $C_{6v}$.
\begin{table}[h!b!p!]
\begin{center}
\begin{tabular*}{0.75\textwidth}{@{\extracolsep{\fill}} |c|  c|  c|  c| }
\hline
\multicolumn{2}{ |c|}
{basis} & $C_{6v}$ & $C_{3v}$ \\
\hline
$x^{2}+y^{2}$, $z^{2}$ & z & $A_{1}$ ($\Gamma_{1}$) & $A_{1}$ ($\Gamma_{1}$) \\
\hline
 & $R_{z}$ & $A_{2}$ ($\Gamma_{2}$) & $A_{2}$ ($\Gamma_{2}$) \\
\hline
$(xz,yz)$ & $(x,y)$,$(R_{x},R_{y})$ & $E_{1}$ ($\Gamma_{5}$) & \\
\cline{1-3}
$(x^{2}-y^{2},xy)$ &   & $E_{2}$ ($\Gamma_{6}$) & \raisebox{1.3 ex}[0pt]{$E$ ($\Gamma_{3}$)} \\ 
\hline
\end{tabular*}
\caption{ Basis table of $C_{6v}$ and $C_{3v}$ and their compatibility relations. $R_{x}=yp_{z}-zp_{y}$ is the angular momentum component around axis $x$, the others are obtained by cyclic permutation.}
\end{center}
\end{table}
%
%
%
%
%
%
Explicitly we consider a 3d magnetic element, and ignore the crystal electric field splitting so that the energy levels for different reps are still degenerate or quasi-degenerate. The wave function should be constructed according to the symmetry group of $C_{6v}$. The reason is that the nonvanishing hybridization should be invariant under the operations in $C_{6v}$ which is obtained as $\Gamma_{i}\otimes\Gamma_{i}$ by group theory. From Table 2, we infer $d_{z^{2}}$ belongs to $\Gamma_{1}$, ($d_{xz}$,$d_{yz}$)  to $\Gamma_{5} (E_{1})$, and  ($d_{x^{2}-y^{2}}$,$d_{xy}$)  to $\Gamma_{6} (E_{2})$. These 3d orbitals are well defined. Having specified the states of the impurity, we consider  the basis of $C_{6v}$ reps by combining the basis of $E$ in $C_{3v}$. We find
\begin{eqnarray}
\phi^{\Gamma_{5}}_{\mathbf{k}\sigma} &=& (1/\sqrt{2})(\phi^{A}_{K_{+}\mathbf{k}\sigma}-\omega\phi^{B}_{K_{-}\mathbf{k}\sigma}, \phi^{A}_{K_{-}\mathbf{k}\sigma}-\omega^{*}\phi^{B}_{K_{+}\mathbf{k}\sigma} )^{T}, \notag \\
\phi^{\Gamma_{6}}_{\mathbf{k}\sigma} &=& (1/\sqrt{2})(\phi^{A}_{K_{+}\mathbf{k}\sigma}+\omega\phi^{B}_{K_{-}\mathbf{k}\sigma}, \phi^{A}_{K_{-}\mathbf{k}\sigma}+\omega^{*}\phi^{B}_{K_{+}\mathbf{k}\sigma})^{T},
\label{g5g6}
\end{eqnarray}
where $\omega=e^{i2\pi/3}$. Therefore, the hybridization Hamiltonian for 3d magnetic impurity is
\begin{equation}
H_{\text{hyb}}=\sum_{\alpha\mathbf{k}\sigma}\left(v^{\alpha}_{\mathbf{k}}c^{\dagger}_{\alpha\mathbf{k}\sigma}f_{\alpha\sigma}+h.c.\right),
\label{hybh}
\end{equation}
where $\alpha=\Gamma_{1},\Gamma_{5},\Gamma_{6}$ reps, and the latter two are 2D reps.
The creation and annihilation operators are obtained by defining the field operators based on the respective basis of  the corresponding rep in the standard way.
This situation resembles the Coqblin - Schrieffer model \cite{csmodel}, however the Hamiltonian now is written in terms of the irreducible reps of the systems.  
\\
\textit{Kondo temperature under large U.-} We calculated the Kondo
temperature for the Anderson model in  ADA case using the equation
of motion method \cite{lacoix} that  gives consistent results for
the susceptibility with those derived by other methods at high
temperature \cite{kash} and delivers  the correct Kondo
temperature \cite{luo}.
The Kondo temperature for  large $U$  is  proportional to
$\exp{\frac{\pi(\varepsilon_{0}-\mu)}{\Delta|\mu|}}$, where
$\Delta=\frac{\pi v^{2}_{0}}{D^{2}}$, $D$ is the energy cutoff.
If $\varepsilon<\mu$ and $\mu\rightarrow 0$, the Kondo
temperature tends to zero. The impurity is then decoupled
from graphene. If $\varepsilon<\mu$ and with increasing $|\mu|$
the Kondo temperature  increases exponentially. Thus, a
small gate voltage may change the Kondo temperature dramatically
\cite{sengupta}.

\section{Concluding remarks}
Summarizing, we investigated
the Kondo model for a magnetic impurity in graphene in an adatom configuration.
 Starting from the tight-binding Anderson model we applied
a partial wave method to expose the coupling flavors. 
The Kondo effect in graphene behaves as follows:
\emph{i)} For ADA case, only one flavor  couples to the impurity. \emph{ii)} For
the ADC case, the number of channels is determined by the multiplets of the impurity
as well. Multi-flavor character shows up in this case.  \emph{iii) }  Based on the irreducible basis of the reps of the system, A Coqblin-Schrieffer-like hybridization model is derived for a 3d magnetic atom. $d_{z^{2}}$, ($d_{xz}$,$d_{yz}$), and ($d_{x^{2}-y^{2}}$,$d_{xy}$) couple with the $\Gamma_{1}$,  $\Gamma_{5} (E_{1})$, and $\Gamma_{6} (E_{2})$ reps respectively. The bases for these reps of graphene are derived explicitly.
\emph{iiii)} The degeneracy of the two Dirac points only
leads to a higher $T_K$   for  ADA case.
Applying a gate voltage the exchange coefficients are non-zero
due to a finite DOS at $E_F$ and the Kondo effect should be observable.

We thank  C.-L. Jia, M.A.N. Ara\'ujo and A. Rosch for useful discussions. The work is supported by the excellence cluster
"Nanostructured Materials" of the state Saxony-Anhalt.

%

\end{document}